\documentclass[%
 reprint,
 amsmath,amssymb,
 aps,
]{revtex4-1}

\usepackage{graphicx}
\usepackage{dcolumn}
\usepackage{bm}

\begin{document}

\preprint{APS/123-QED}

\title{Measuring the orbital angular momentum of electron vortex beams using a forked grating}

\author{Koh Saitoh}
 \email{saitoh@esi.nagoya-u.ac.jp}
 \affiliation{EcoTopia Science Institute, Nagoya University, Nagoya 464-8603, Japan}

\author{Nobuo Tanaka}
 \affiliation{EcoTopia Science Institute, Nagoya University, Nagoya 464-8603, Japan}
  
\author{Yuya Hasegawa}%
 \affiliation{Department of Crystalline Materials Science, Nagoya University, Nagoya 464-8603, Japan}

\author{Kazuma Hirakawa}%
 \affiliation{Department of Crystalline Materials Science, Nagoya University, Nagoya 464-8603, Japan}


\author{Masaya Uchida}
\affiliation{
 Advanced Science Research Laboratory, Saitama Institute of Technology, Fukaya 369-0293, Japan
}%


\date{\today}

\begin{abstract}
The present study experimentally examines how an electron vortex beam with orbital angular momentum (OAM) undergoes diffraction through a forked grating. 
The $n$th-order diffracted electron vortex beam after passing through a forked grating with a Burgers vector of 1 shows an OAM transfer of $n\hbar$. 
Hence, the diffraction patterns become mirror asymmetric owing to the size difference between the electron beams. 
Such a forked grating, when used in combination with a pinhole located at the diffraction plane, could act as an analyzer to measure the OAM of input electrons. 

\begin{description}

\item[PACS numbers]
41.85.-p, 42.40.Eq, 42.50.Tx

\end{description}
\end{abstract}

\pacs{Valid PACS appear here}
\maketitle

The discovery of the electron vortex beam, $i.e.$, 
electrons propagating through free space with quantized orbital angular momentum (OAM) \cite{uchida10}, 
attracted a substantial amount of attention 
owing to the unique physical properties of the beams and their potential applications to new electron microscopy and spectroscopy \cite{verbeeck10,mcmorran11}. 
Electron vortex beams with OAM are easily generated in conventional microscopes by a phase plate \cite{uchida10}, 
a forked grating \cite{verbeeck10,mcmorran11}, and spiral zone plates \cite{verbeeck11,saitoh12JEM}. 
Furthermore, McMorran $et$ $al$. \cite{mcmorran11} and Saitoh $et$ $al$. \cite{saitoh12JEM} independently reported 
that it is possible to generate electron vortex beams with OAM as high as 100$\hbar$ and 90$\hbar$.

Both angular and linear momentum can be transferred via various scattering processes between an incident beam and a scatterer, such as an atom and a solid. 
In both elastic and inelastic scattering processes, 
linear and angular momentum are conserved in a closed system. 
The angular-resolved electron energy-loss spectroscopy (EELS) technique is a prominent application of the conservation rule of linear momentum; 
a particular excited state of the atom can be probed by analyzing the angular dependence of the scattered waves \cite{egerton_eelsbook,leapman83,botton05,saitoh06,saitoh12JAP}. 
Electron OAM can be thought of as new degrees of freedom for free electrons, 
as in the case of photon OAM \cite{allen92,allen99}. 
For example, one expects that measuring scattered electron OAM would allow probing of electronic states with particular OAM quantities in an atom \cite{babiker02,lloyd12}. 
In this context, techniques to measure the OAM of free electrons are needed for applications such as scattering and spectroscopy experiments. 
In the present letter, we seek to explore the measurement of electron OAM by using a nano-fabricated, forked grating. 
We also seek to explain the results of recent experimental studies by Verbeeck $et$ $al$., 
in which the observation of dichroism is reported in EELS of ferromagnetic Fe thin films using electron vortex beams \cite{verbeeck10}. 

\begin{figure}
\includegraphics[width=8.6cm]{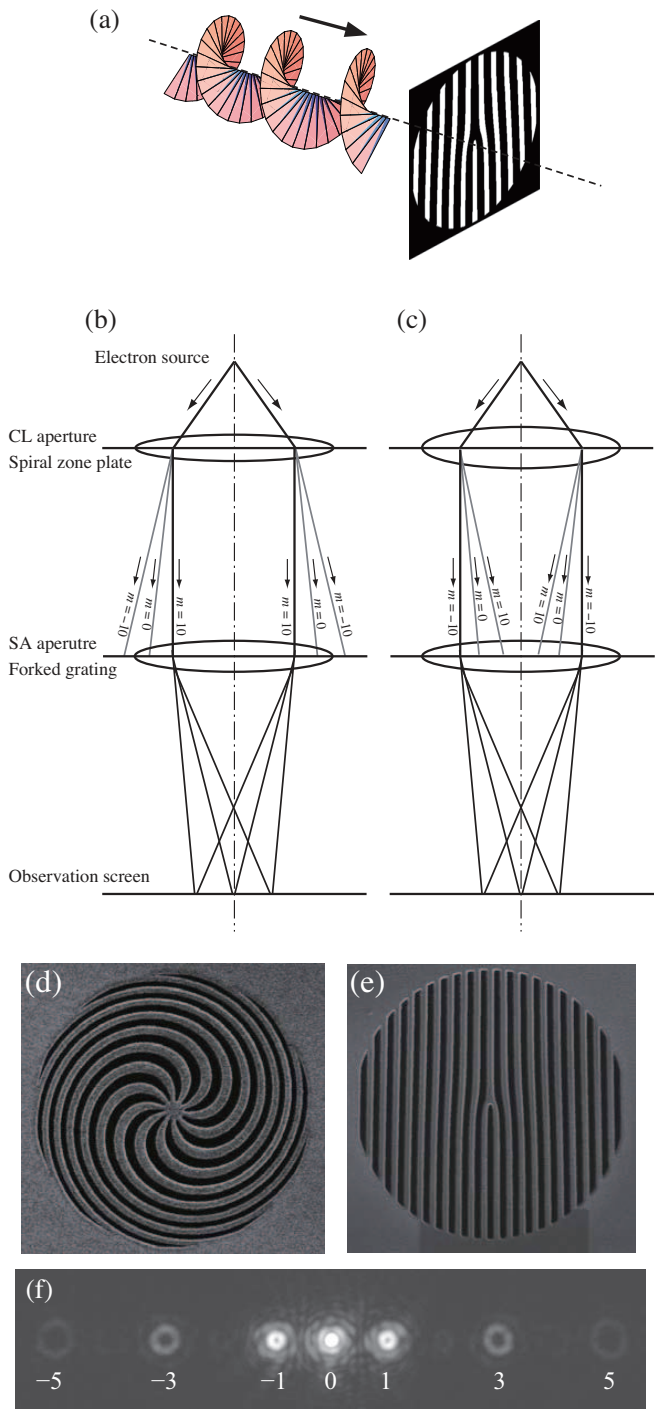}
\caption{\label{fig1} (a) Schematic drawing of the present experiment. (b),(c) Raypath diagrams of the present experimental setups. (d) Spiral zone plate with a diameter of about 20 $\mu$m introduced to the condenser lens aperture position. (e) Forked grating with a diameter of about 30 $\mu$m introduced to the selected-area aperture position. Electron vortex beams of $m$ = 10 (b) and $m$ = $-$10 (c) are set to satisfy the parallel beam condition and are diffracted by the forked grating, forming a diffraction pattern at the screen. (f) Diffraction pattern created from the forked grating shown in (e).}
\end{figure}

A series of diffracted electron vortex beams with quantized OAM are formed by forked gratings \cite{verbeeck10,mcmorran11} and spiral zone plates \cite{verbeeck11,saitoh12JEM}. 
In the present experiment, we inject electron vortex beams into a forked grating and observe how the output beams propagate (Fig.~\ref{fig1}(a)). 
Figures \ref{fig1}(b) and \ref{fig1}(c) show a schematic diagram of the experimental setup of the present study.
The binary masks of the spiral zone plates (Fig.~\ref{fig1}(d)) and the forked gratings (Fig.~\ref{fig1}(e)) were fabricated from Si$_{3}$N$_{4}$ membranes with a thickness of 50 nm, 
on which PtPd films with a thickness of about 100 nm were deposited on each side of the membrane using a focused-ion-beam instrument (Hitachi FB-2100). 
The spiral zone plates and forked gratings were inserted into the condenser lens aperture position and selected-area aperture position, respectively, of a transmission electron microscope (JEOL JEM-2100F), 
which was operated at an acceleration voltage of 200 keV. 
The diffraction patterns were recorded at a camera length of 100 cm by a 16-bit CCD camera with 2 k × 2 k pixels, mounted at the end of a Gatan imaging filter. 
This imaging filter is effective for observing detailed features of the patterns because of its intrinsic high magnification. 

A spiral zone plate with a topological charge of 10 (Fig.~\ref{fig1}(d)) was inserted at the condenser lens system. The spiral zone plate produces a series of electron vortex beams with different topological charges, which focus into or diverge from points located at different positions along the propagation direction \cite{saitoh12JEM}. 
The $n$th-order electron beam that is produced by the present spiral zone plate has an OAM of $10n\hbar$. 
The convergence angles of the electron vortex beams can be adjusted by changing the excitation of the condenser lens system. Here the 1st- or $-$1st-order electron vortex beam was set to satisfy the parallel illumination condition onto the forked grating. 
It should be noted that when the 1st-order electron vortex beam satisfies the parallel illumination condition, the $-$1st-order beam does not, and vice-versa. 
A grating with a fork-like dislocation with a Burgers vector of 1 (Fig.~\ref{fig1}(e)) was inserted at the selected-area aperture position. 

Figure \ref{fig1}(f) shows a diffraction pattern created by a forked grating that is illuminated by a plane wave with zero OAM. The pattern shows a sharp spot (where $m$ = 0) at the center as a transmitted beam, and ring-shaped peaks with OAM of $\pm\hbar$ are observed at either side of the transmitted electron beam. 

\begin{figure*}
\includegraphics[width=15cm]{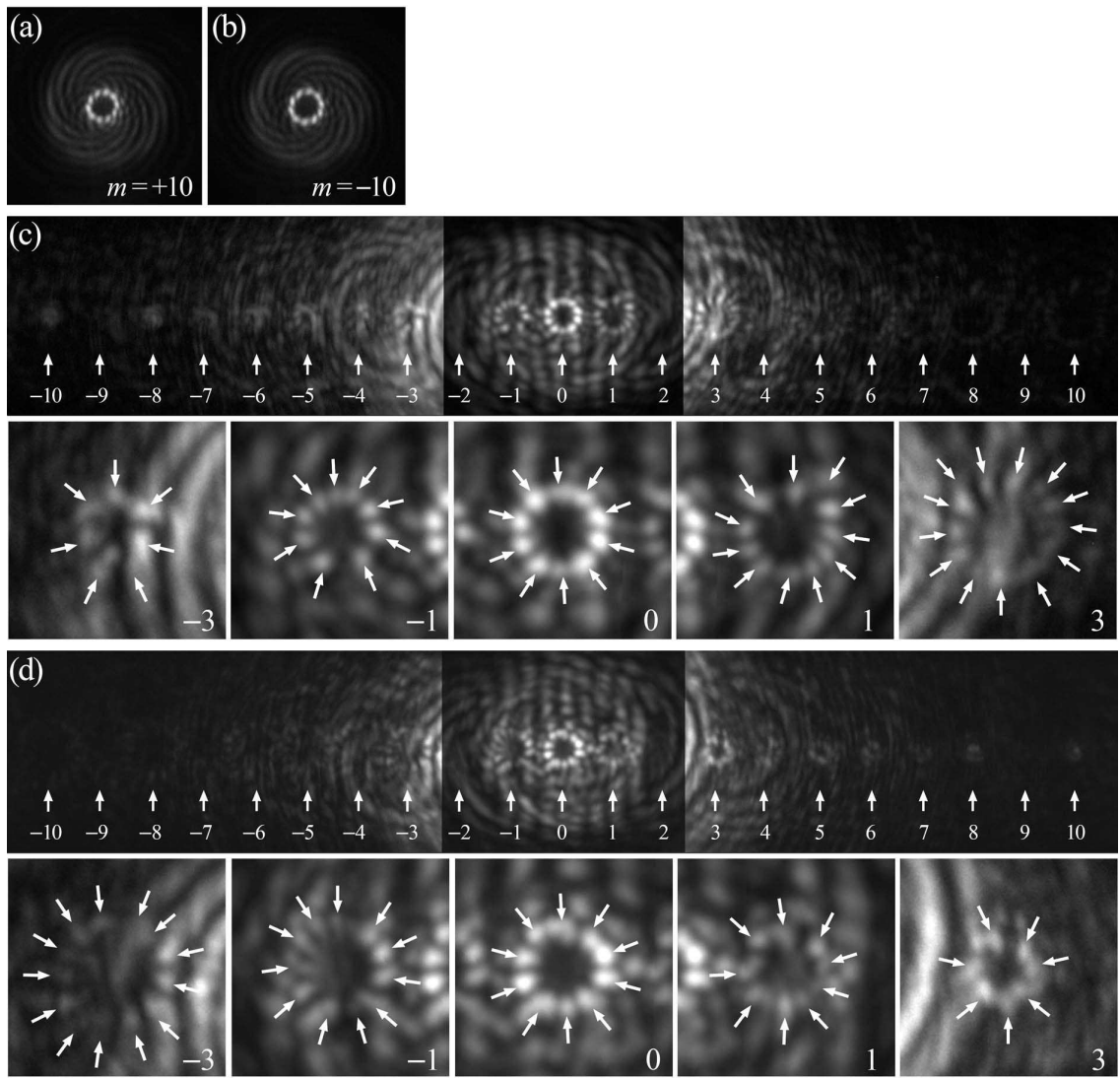}
\caption{\label{fig2} Incident electron vortex beams with $m$ = +10 (a) and $m$ = $-$10 (b), and diffraction patterns of the vortex beams with $m$ = 10 (c) and $m$ = $-$10 (d) generated by the forked grating shown in Fig. 1(e). A series of diffracted rings are aligned in the horizontal direction. In the case of an incident beam where $m$ = 10, the diameters of the diffracted rings increase with the diffraction order, whereas in the case where $m$ = $-$10, the diameters decrease with diffraction order.}
\end{figure*}

Figures \ref{fig2}(a) and \ref{fig2}(b) show electron vortex beams with OAMs of 10$\hbar$ and $-$10$\hbar$, respectively, produced by the spiral zone plate. 
Each of the electron beams show a ring composed of 10 peaks at the center. The electron OAM can be determined by the number of peaks in the ring when the beam is produced by a spiral zone plate \cite{saitoh12JEM}.

We then investigated how an electron vortex beam with a particular OAM quantity undergoes diffraction in the forked grating. 
Figure \ref{fig2}(c) shows an electron diffraction pattern for an incident electron vortex beam with m = 10$\hbar$. 
Here the incident electron beam can be set to satisfy the parallel illumination condition with the forked grating by adjusting the excitation of the condenser lens system. 
The diffraction pattern shows a series of diffracted rings, aligned in the horizontal direction, as indicated by the arrows. The central ring, composed of 10 peaks, is the transmitted beam with $m = 10\hbar$. 
The 1st- and $-$1st-order diffracted electron beams show similar ring-like features, 
but have 11 and 9 peaks with larger and smaller diameters than the transmitted ring, respectively. 
This indicates that the electron OAMs of the 1st- and $-$1st-order diffracted beams are 11$\hbar$ and 9$\hbar$, respectively. 
The 3rd- and $-$3rd-order diffracted beams, denoted as 3 and $-$3 in Fig. 2(a), show faint ring features having even larger and smaller diameters than the 1st- and $-$1st-order rings, respectively. 
The 2nd- and $-$2nd-order diffracted beams are not observed because of destructive interference for all even orders. Up to 10th- and $-$10th-order diffracted beams are observed in the present experiment. 
Fringes surrounding the diffracted rings are caused by the transmitted beam through the spiral zone plate, 
which are not focused at the observation plane. 
The fringes are overlaid, interference with the diffracted rings occurs, 
and the peak intensities in the rings are modulated. 
Therefore, some of the peaks in the rings are not clearly observed and the original rotational symmetry of the rings is broken.

The diffraction pattern for a vortex beam corresponding to $m = -10\hbar$ is shown in Fig.~\ref{fig2}(d). 
The observed pattern shows a series of diffracted rings aligned in the horizontal direction, as in Fig.~\ref{fig2}(c). 
However, the diffraction pattern is horizontally inverted from that shown in Fig.~\ref{fig2}(c). 
The transmitted (0th-), 1st-, and $-$1st-order diffracted rings show 10, 9, and 11 peaks, respectively, 
indicating that the electron OAM of the 1st- and $-$1st-order beams are $-9\hbar$  and $-11\hbar$, respectively.
Our experimental results indicate that the forked grating with a Burgers vector of $b$ = 1 transfers not only linear momentum but also OAM, where the electron OAM transfer of the $n$th-order diffracted electron beam is $n\hbar$\cite{mair01,leach02,moreno09,bekshaev10}.

\begin{figure}
\includegraphics[width=8cm]{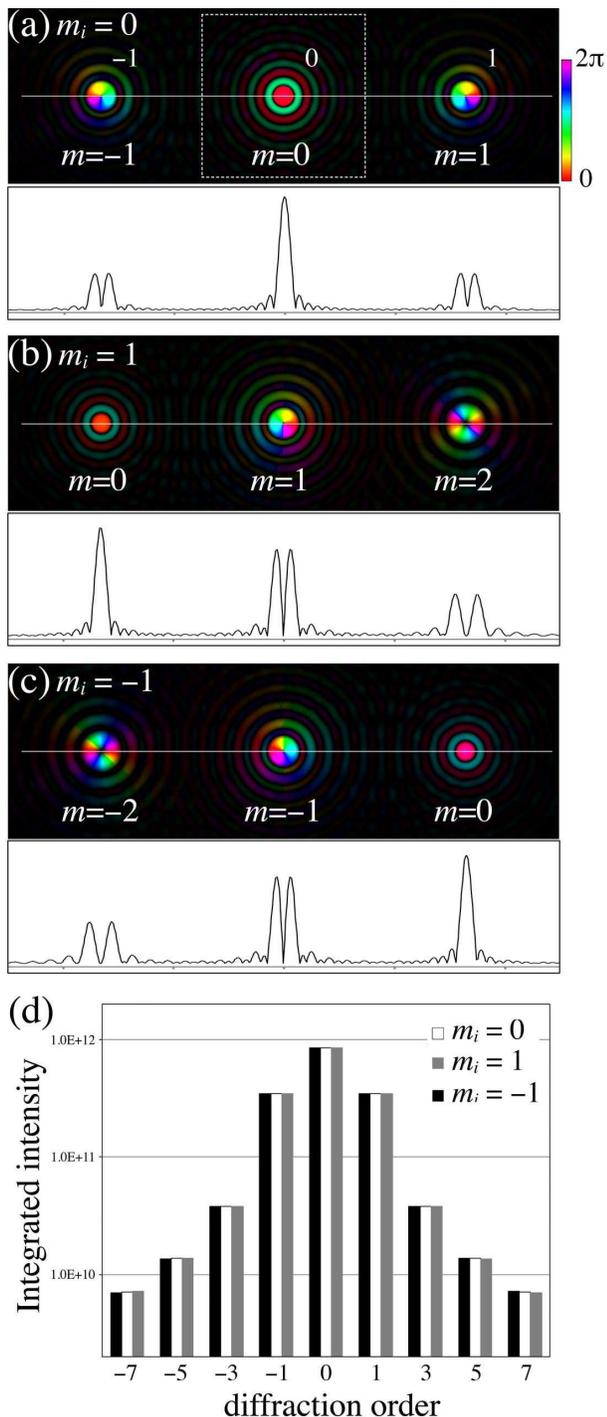}
\caption{\label{fig3} Simulated diffraction patterns of the forked grating used in the present experiment, where OAMs of the incident beams $m_{i}$ are set to be $m_{i}$ = 0 (a), +1 (b), and $-$1 (c). 
Color coding indicates the phase distribution from 0 (red) to 2$\pi$ (purple). 
The intensity profile along the line passing through the centers of the diffraction peaks is symmetric for $m_{i}$ = 0, but is asymmetric for $m_{i}$ = +1 and $-$1. 
(d) Integrated intensities of the diffracted beams for $m$ = 0, +1, and $-$1. One of the integration areas is indicated by the dotted white squares in (a). 
Note that the integrated intensities of the $n$th and $-n$th diffracted beams are almost identical, regardless of the OAM of the incident beam. }
\end{figure}

As a check on our experimental results, we carried out a simulation study based on Fresnel propagation theory \cite{born&wolf}. 
Figures \ref{fig3}(a), \ref{fig3}(b), and \ref{fig3}(c) show simulated diffraction patterns from the forked grating that was used in the present experiment (Fig.~\ref{fig1}(b)), 
in which the OAMs of the incident beams, $m_{i}$, are assumed to be 0, $+\hbar$, and $-\hbar$, respectively.
Color coding indicates the phase distribution from 0 (red) to 2$\pi$ (purple).
The line profile shown in the lower part of each figure is an intensity profile along the horizontal line passing through the centers of the diffraction peaks, 
indicated by the white lines. 
We can see that the diffraction patterns are mirror asymmetric, 
as observed in the present experiment. 
Here we calculated integrated intensities of the diffraction peaks of $\pm1$, $\pm3$, $\pm5$, and $\pm7$ for the incident beams of $m_{i}$ = 0, $+\hbar$, and $-\hbar$ (Fig.~\ref{fig3}(d)). 
Importantly, regardless of the OAM of the incident beam, 
the total intensities of the $n$th- and $-n$th-order diffracted beams formed by the forked grating are almost identical. 
Indeed, the difference in the integrated intensities of the 1st- and $-$1st-order beams is about 0.2\% in the case of $m_{i} = \pm\hbar$. 
This result suggests that it is difficult to measure the OAM by simply comparing the total intensities of the $\pm$1st-order diffracted beams. 
The difference between the $n$th- and $-n$th-order diffracted intensities increases as $n$ grows larger. 
The difference reaches up to 2\% by taking the 7th- and $-$7th-order diffracted beams. 
Such small differences can be enhanced by reducing the integration area, as is expected from the asymmetric diffraction patterns. 
The largest differences are obtained by taking only the central positions of the diffracted beams because the diffracted vortex beam with non-zero OAM has zero intensity at the beam center and the diffracted beam with zero OAM has non-zero intensity at the center. 

A forked grating that transfers electron OAM becomes an effective tool for the OAM measurement of free electrons when used in combination with a pinhole. 
This technique is reminiscent of the single-mode fiber in optics \cite{mair01}. 
Assuming that an electron vortex beam with an OAM of $m\hbar$  is incident upon a forked grating with $b = 1$, 
the $m$th-order diffracted beam is a plane wave with a sharp peak corresponding to an electron OAM transfer of  $m\hbar$. 
Furthermore, if we place a pinhole at the position of the $m$th diffraction peak on the diffraction plane, 
it selects only sharp diffraction peaks corresponding to electrons with zero OAM. 
That is, the forked grating combined with the pinhole could act as an analyzer (sorter) to measure the OAM of input electrons.

Such an electron OAM analyzer has a broad range of applications, most notably in materials science. 
In a particular inelastic scattering process accompanied by inner-shell excitations of an atom, 
a momentum transfer between the scattered free electron and the excited atomic electron can be anisotropic, 
reflecting an anisotropy of the initial and/or final states of the excited atomic electron \cite{egerton_eelsbook,leapman83,botton05,saitoh06,saitoh12JAP}. 
The excitation from the initial state to the final state is regarded to a good approximation as a dipole transition, which accompanies an angular momentum transfer of $\Delta l = \pm 1$. 
Therefore, OAM can be transferred to the free electron via inelastic scattering during the recoil of the transition of the atomic electron. 
A forked grating with $b = \pm 1$, which is set at a post-specimen plane, 
can be used to distinguish such inelastically scattered electrons with OAM of $\hbar$ or $-\hbar$ by observing the 1st- and $-$1st-order diffracted beams. 
In addition, one can use a pinhole to select only electrons with zero OAM (or another particular OAM quantity). 
Because magnetic spin is coupled to the final state of the excitation, 
that is, spin-up and spin-down states are correlated to the transitions of $\Delta l$ = +1 and $-$1, respectively, 
the magnetic spin can be probed by measuring the OAM of the inelastically scattered electron using the forked grating. 
Furthermore, we note that one can apply the electron OAM analyzer for not only dipole interactions but also quadrupole and higher-order multipole interactions.

The present results provide an important clue to understand the recently observed dichroism by Verbeeck $et$ $al$. \cite{verbeeck10}. 
They affixed the forked grating posterior to a Fe thin film and observed a significant difference in EELS signals of the Fe-$L_{23}$ peaks between the 1st- and $-$1st-order diffracted electron beams of the forked grating. 
If the inelastically scattered electrons leaving the Fe film form a vortex beam where $m = \hbar$, 
the 1st- and $-$1st-order diffracted beams correspond to $m = 2\hbar$ and $m = 0$, respectively; 
on the other hand, if the inelastically scattered electrons leaving the Fe film form a vortex beam 
where $m = -\hbar$, the 1st- and $-$1st-order beams correspond to $m = 0$ and $-2\hbar$ , respectively. 
As shown in Fig.~\ref{fig3}, the diffraction patterns in both cases will be mirror asymmetric. 
However, for the above mentioned reasons, the dichroism in the EELS signals cannot be explained simply by comparing the total intensities of the entire $\pm$1st-order diffracted beams, even if there are differences in the probability of a transition occurring between $\Delta l$ = +1 and $-$1. 
One might need to consider ``pinhole effects" for selecting the central part of the ``sharp" diffracted beams in EELS experiments. 

In conclusion, we investigated the OAM transfer of electron vortex beams by a forked grating. 
The $n$th-order diffracted electron vortex beam generated by a forked grating with a Burgers vector of 1 showed an OAM transfer of $n\hbar$. 
Such a forked grating, when used in combination with a pinhole, could be used as an electron OAM analyzer. 
Our results could lead to new approaches in electron microscopy and spectroscopy. 
The present method can be applied not only to magnetic materials, but also to nonmagnetic materials. 
For instance, it is interesting to measure the OAM of secondary electrons in electron microscopy and photoelectrons in photoelectron spectroscopy. 
Furthermore, it could also be applied to the measurement of electric and magnetic fields because non-uniform electric and magnetic fields would lead to an OAM transfer.

\begin{acknowledgments}
The present work was partially supported by Grant-in-Aid for Scientific Research (A) (No. 23241036), the Ministry of Education, Culture, Sports, Science and Technology, Japan, and the Mitsubishi Foundation.
\end{acknowledgments}

\bibliography{LA13591}

\end{document}